# Artificial Synapse with Mnemonic Functionality using GSST-based Photonic Integrated Memory

**Mario Miscuglio[1], Jiawei Meng[1], Omer Yesiliurt[2], Yifei Zhang[3], Ludmila J. Prokopeva[2], Armin Mehrabian[1], Juejun Hu[3], Alexander V. Kildishev[2] and Volker J. Sorger[1]***

[1] Department of Electrical and Computer Engineering
George Washington University, 800 22nd Street NW, Washington, DC 20052, USA

[2] Birck Nanotechnology Center,
School of ECE, Purdue University, West Lafayette, IN 47907, USA

[3] Department of Materials Science & Engineering
Massachusetts Institute of Technology, Cambridge, MA, USA

*Corresponding Author, sorger@gwu.edu

*Abstract ─* Machine-learning tasks performed by neural networks demonstrated useful capabilities for producing reliable, and repeatable intelligent decisions. Integrated photonics, leveraging both component miniaturization and the wave-nature of the signals, can potentially outperform electronics architectures when performing inference tasks. However, the missing photon-photon force challenges non-volatile photonic device-functionality required for efficient neural networks. Here we present a novel concept and optimization of multi-level discrete-state non-volatile photonic integrated memory based on an ultra-compact (<4μm) hybrid phase change material GSST-silicon Mach Zehnder modulator, with low insertion losses (3dB), to serve as node in a photonic neural network. An optimized electro-thermal switching mechanism, induced by Joule heating through tungsten contacts, is engineered. This operation allows to change the phase of the GSST film thus providing weight updating functionality to the network. We show that a 5 V pulse-train (<1 μs, 20 pulses) applied to a serpentine contact produces crystallization and a single pulse of longer duration (2 μs) amorphization, used to set the analog synaptic weights of a neuron. Emulating an opportuney trained 100×100 fully connected multilayered perceptron neural network with this weighting functionality embedded as photonic memory, shows up to 93% inference accuracy and robustness towards noise when performing predictions of unseen data.

*Index Terms ─* Phase change materials, PCM, Photonic Memory, GSST, Neuromorphic, Integrated Photonics, Neural Network, non-volatile.

## I. INTRODUCTION

The past decades have been marked by an exponential increase in demand for high-speed and energy-efficient computer architectures. Moore's law and Dennard-scaling have reached their respective limits [1]; therefore, microelectronics faces fundamental trade-offs in terms of power efficiency when processing large data volumes and complex systems with short delay. Some of these computational or processing challenges could be addressed with neural networks (NN); these networks comprised of neurons, loosely modeled based on their biological counterparts, take up, processes, and transmit information through electrical signals. Their main operations are weighted additions of the input signals (multiply and accumulate, MAC) and a nonlinear activation function (NLAF), i.e. threshold. The training phase of a NN consists of a feeding large dataset to the network and by back-propagation recursively adjust the weights for modelling the data. After being trained a faster and more efficient version of a NN can infer and perform prediction tasks on new (unseen) data. Due to this large amount of recursive and iterative operations, NNs are usually implemented in Tensor Process Units (TPUs) and Graphic Process Units (GPUs), which are optimized architectures that parallely and efficiently perform the dot-product multiplication, summation, and nonlinear (NL) thresholding on the input data, providing high efficiency and ultra-high throughput [2] for specific tasks. Thanks to dedicated modules, they perform inference tasks more efficient than general-purpose CPU equivalents, however, these approaches still rely on electronic transport and are bound by the speed and power limits of the interconnects inside the circuits hence affected by RC parasitic effects.

In contrast, to electronics, integrated photonics can provide low delay interconnectivity which meets the requirements for node-distributed non-Von Neumann architectures that can implement NNs relying on dense node-to-node communication. Moreover, in trained networks the weights found during training are fixed and are only sporadically updated, therefore weighted addition (MAC) and vector matrix multiplication (VMM) can be effortlessly and passively performed in photonics, by means of phase modulation or wavelength division multiplexing (WDM) using networks of linear EO modulators, MZMs [3], [4] or micro-ring modulators [5], [6]. Thus, once the NNs weights are SET, after training, the delay in the network is given by the time-of-flight of the photon, which for large network is in the 10's ps range, plus the back-end O-E conversion by the photodetector (<100 ps).

However, the functionality of memory for storing the trained weights is not straightforwardly achieved in optics[7], [8], or at least in its non-volatile implementation, and therefore requires additional circuitry and components (i.e. DAC, memory) and related consumption of static power, sinking the overall benefits (energy efficiency and speed) of photonics. Therefore, computing AI-systems and machine-learning (ML) tasks, while transferring and storing data exclusively in the optical domain, is highly desirable because of the inherently large bandwidth, low residual crosstalk, and short-delay of optical information transfer.[9] The non-volatile retention of information in integrated photonics can be provided by the light-matter interaction in phase change memory (PCM)[10]–[14].

Germanium-antimony-tellurium (GST) is a PCM from the group of chalcogenide glasses and is characterized by a remarkable variation of the complex refractive index between its crystalline and amorphous states. In such a material, the variation of the phase can be induced by local heating, either thermally (heaters), electrostatically (contacts), or all-optically (laser). Recently, GST have been also employed in photonic NNs [15]–[17]. However, even if these materials exhibit large contrast of both refractive index ($\Delta n$) and optical loss ($\Delta k$), simultaneously, they are characterized by relatively high insertion losses, high switching energy and poor number of cycles, which can potentially limit the number of neurons and the depth of the network, and concurrently the number of times that the network can be updated. Due to these limitations, new classes of specialized optical PCMs are investigated.

Here, we leverage on a recently engineered class of optical PCMs, based on Ge–Sb–Se–Te (GSST) alloy[18], whose amorphous state is not characterized by high absorption coefficient, and upon phase change, its refractive index is still subjected to unitary modulation. The optimized alloy, $Ge_2Sb_2Se_4Te_1$, combines broadband transparency (1–18.5 μm), large optical contrast ($\Delta n = 2.0$), and significantly improved glass forming ability. Thence, we design a low loss non-volatile photonic memory, using a balanced GSST-based Mach Zehnder Modulator and develop a numerical framework for optimizing the heaters configuration and evaluating the temporal switching response of such GSST-based photonic memory.

Exploiting the low delay interconnectivity of photonic integrated chips and the non-volatile transitions of PCM, an all-optical (AO) trained NN, that effortlessly performs dot-product functionality can be achieved, enabling intelligent computing functionality at the time-of-flight of the photon.

## II. RESULTS

### A. Optical constants of phase change materials

Improvements in the field of non-volatile photonic memory pertains the engineering of processes aimed at synthetizing more effective and efficient films based on phase change alloys, in which a concurrent minimization of the losses and maximization of the modulation between amorphous and crystalline state allows to keep information longer in the optical-domain, i.e. avoids cumbersome O-to-E-to-O conversions[13], [19].

This, however, requires a wisely engineered material process, e.g. interfacial PCM (GeTe/Sb$_2$Te$_3$)[13] and optimized alloys[18]. Contrary to regularly used GST (Fig. 1a), GSS$_4$T$_1$ exhibits a 3 orders of magnitude lower absorption coefficient while preserving a large $\Delta n$ of 2.1 to 1.7 across the near- to mid-IR bands, suggesting its use as photonic memory in an electro-refractive scheme, such as MZI or ring-based cavities. Additionally, the GSS$_4$T$_1$ film is characterized by a high index ratio, $\Delta n/\kappa = 5$ (also often used as figure of merit for nonlinear materials), due to its low index contrast and relatively small $\kappa$, even without any metal-insulator transition. Electro-refractive photonic memory based on hybrid GSS$_4$T$_1$-Silicon waveguide operating at 1550 nm would allow to fabricate photonic memory devices with remarkably large modulation dynamic and contained losses (**Table. 1**).

In this paper, we use the tabulated ellipsometry data for modeling. However, for fully coupled multiphysics modeling including thermal processes and switching, it is possible to develop a time-domain multivariate model with temperature and crystallization level as parameters. Normally, data fitting for PCMs is done in the frequency domain with the Code-Lorentz, Tauc-Lorentz, and Gauss models for each dataset, that do not allow dependence on parameters and straightforward time-domain implementation. The full multivariate time-domain model however can be built using the generalized dispersive material model[20].

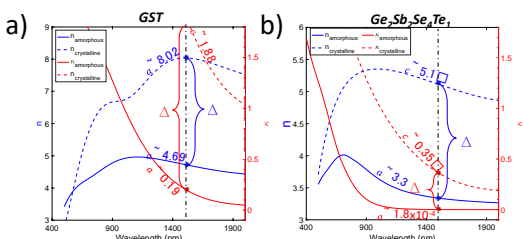

**Figure 1.** Experimentally obtained (ellipsometry) optical properties of phase change material (PCM) films: GST (a), and Ge₂Sb₂Se₄Te₁(b). Real ($n$, left y-axis) and imaginary ($\kappa$, right y-axis) parts of the refractive indices of the amorphous (solid line) and crystalline alloys. (dashed line). The GSST (right) shows a strong unity $\Delta n$, while simultaneously showing small induced loss, $\Delta\kappa = 0.4$ making it a promising candidate for non-volatile phase-shifting photonic devices such as modulators, tunable structures, or directional-coupler-based 2x2 switches.

**Table 1:** Complex refractive index ($n+i\kappa$) of different GST and GSST (Ge-Sb-Se-Te) materials at 1550 nm, characterized by ellipsometry. For our study, we consider **GSS₄T₁** (values taken from [21])which displays a particularly high Figure of Merit, defined as $\Delta n/\kappa$, (5.02).

| Material | AMORPHOUS | | CRYSTALLINE | | FOM |
|---|---|---|---|---|---|
| | **n** | **κ** | **n** | **κ** | |
| **GST225** | 4.690 | 0.192 | 8.027 | 1.882 | 1.774 |
| **GSS1T4** | 4.725 | 0.208 | 7.704 | 1.464 | 2.035 |
| **GSS2T3** | 4.800 | 0.220 | 7.059 | 1.444 | 1.565 |
| **GSS3T2** | 4.192 | 0.056 | 6.800 | 1.049 | 2.485 |
| **→ GSS4T1** | 3.325 | 1.8x10⁻⁴ | 5.083 | 0.350 | **→ 5.02** |

## B. Mode analysis

With photonic networks on chip in mind [22], when the network is trained offline using emulators, such as TensorFlow, which accounts for the physics of the devices and their functionalities[6], [4], the extracted weights are set by selectively 'writing' portions of the GSST deposited on the waveguides, by heat induced laser irradiation, or local electrostatic heating, which promotes crystallization, and consequently modifies the waveguide modal refractive index in a reversible process (**Fig. 2**).

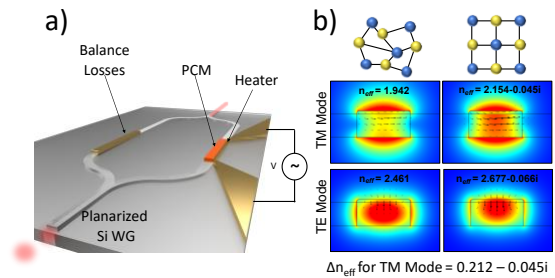

$\Delta n_{eff}$ for TM Mode = 0.212 – 0.045i

**Figure 2.** Schematic of an electro-optic modulator based on a balanced Mach Zehnder interferometer (MZI), such as used to program weights (dot-products) of a photonic NN. b) Fundamental transversal electric (TE) and Transverse magnetic mode profiles (normalized electric field) of the GSST-Silicon hybrid waveguide at 1550 nm for amorphous and crystalline GSST show a strong index (real-part) difference ~ 0.2, while incurring a relatively low loss $\Delta\kappa = 0.4$. Black arrows represent the direction and intensity of the magnetic field (Hₓ,Hᵧ).

To derive the effective modal index and the propagation length of the hybrid GSST-silicon waveguide we use eigenmode analysis. A PCM film of 30 nm of Ge-Sb-Se-Te (GSST) [Ge₂Sb₂Se₄Te₁] is considered to be deposited atop of a planarized waveguide and a phase transition is induced by local heaters (described in **Section II.C**).

The model exploits the experimentally measured optical constants of GSST (**Section II.A**), which are obtained using coupled spectroscopic ellipsometry and transmittance/reflectance measurements from the visible through long-wave infrared. The complex effective refractive index ($n_{eff}$) in the amorphous state enables rather low insertion losses for a strong real-part variation $\Delta n_{eff} = n_{eff,\,a} - n_{eff,\,c}$ approximately equal to 0.2 -0.25 for TM and TE mode, respectively (**Fig. 2b**).

Considering such favorably low insertion losses caused by the low absorption coefficient in the amorphous state and the stark variation of the refractive index in the crystalline state, we proceed to designing a balanced passive push-pull Mach Zehnder modulator (MZM) configuration (**Fig. 2a**), in which on both sides the GSST material are deposited in the amorphous condition (aGSST). To modulate the intensity of the signal at its output, the MZI is purposely unbalanced by thermally writing a portion of the GSST film deposited in the "programmable arm" of the MZI (**Fig. 3b**). For a TM mode, for instance, the length of the active part of the modulator is just 3.8 µm short for achieving a π phase shift, when the entire film on the 'recordable' branch has changed to its crystalline phase. To our knowledge, the device is one order of magnitude smaller than one of the most compact MZM ever reported[23], with positive effects on the electrical capacitance improving both response time and power consumption.

The lateral section of the written part of the material corresponds to a "quantized weight", and assuming a stabile writing resolution of about 500 nm[14], same achieved by optical writing, the total amount of available discrete resolution is given by 8 distinct states (3-bit) which can be further improved or condition on the resolution relaxed by extending the device length by multiple of Lπ (**Fig.3a-b**). Additionally, this is a reversible process, which allows to update the weights after many execution times. Interestingly, this solution is not hindered by insertion losses, which are negligible due to the rather low absorption coefficient of GSST at 1550 nm, and the total losses (~3dB) are mainly caused by the balancing mechanism (in this first analysis straightforwardly obtained achieved by placing a gold contact on the balancing arm). As an interim conclusion, this novel PCM MZM features by a micrometer-compact footprint and low insertion losses (<3dB), enabling the implementation of a deep NN (3 layers) which comprises multiple nodes.

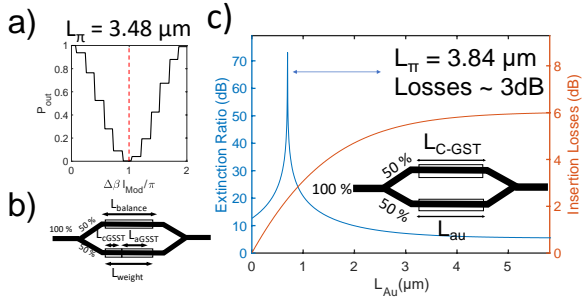

**Figure 2.** Extinction ratio, ER, and insertion loss, IL, performance for the loss balanced MZI. (a) Schematic representation of the balanced GSST MZM. The loss is balanced by depositing additional cGSST on the balancing arm to maximize the ER compensating for the additional losses of the phase changed (cGSST) portion in the programmable arm (a) The normalized output power considering compensated losses, the weighting is quantized since the resolution of the phase changing process is 500nm for altering the active arm of the MZI.

## C. Electrothermal Switching

As mentioned in the previous section, the GSST needs to be locally written according to the weights obtained during the training phase (**Section II.D**). For performing this function, we consider using electro-thermal local switching using heaters. 3D time-dependent multi-physics simulations, including model for heat transfer in solids coupled with the electric currents model, have been carried out.

Preliminary thermal characterization conducted by our group shows that the conductivity of GSST is $0.17\pm0.02$ W/m/K for amorphous phase and $0.43\pm0.04$ W/m/K for crystalline phase, while the heat capacity in amorphous and crystalline phase for GSST film are $1.45\pm0.05$ MJ/m3/K and $1.85\pm0.05$ MJ/m^3/K, respectively. We primarily focus our efforts in optimizing the heaters position with respect to the waveguide to minimize the ohmic losses due to the presence of metal and concurrently lower the threshold voltage for delivering the right amount of heat for inducing a phase transition in the GSST.

The heating elements (**Fig. 5**) considered are tungsten and ITO shaped in a circuitous serpentine (10 µm length, 20 folds), and properly biased they dissipate energy in the form of Joule heat in the surrounding media. In this view, we investigate three different heaters configuration: 1) vertical, using tungsten $W$ (Fig. **6**a) 2) vertical ITO/contact 3) lateral (**Fig. 6b**). Moreover, the temperature at which the GSST reaches the amorphous state is considered in our study around 900 K, whereas for inducing re-crystallization, the GSST needs to be heated above the crystallization temperature (~523 K) but below the melting point, for a critical amount of time, therefore multiple pulses are needed [18].

1) The vertical configuration consists in placing the tungsten ($W$) heating element 250 nm above the silicon waveguide, surrounded by an oxide layer, to minimize the mode-overlap with the metal and eventual scattering, while still in proximity of the GSST layer on top of the waveguide. This configuration requires longer pulse period (6µs) for crystallization and higher threshold voltage (25V) for inducing amorphization/crystallization among the group. The losses introduced by this configuration are extremely low (0.06dB/µm for a propagating TM mode).

2) The lateral configuration consists of two $W$ resistive heaters placed directly in contact with the GSST film, 50 nm away from the waveguide, thus providing more heat to the film locally (lowering the switching threshold), but also storing heat for successive pulses. This configuration even though is more electrothermally efficient is affected by higher insertion losses compared to the previous one (additional ~0.11dB/µm for a propagating TM mode).

3) Being ITO characterized by low optical losses ($n = 1.4, \kappa = 0.2$) at 1550 nm, (electrical resistivity is $0.0016\ \Omega \times$ cm, thermal conductivity is 1340 W/mK), a serpentine was alternatively placed directly on top of the waveguide. ITO heating element provides ~0.14 dB/µm of additional insertion losses and can provide enough heat for crystallization by applying 5-V train and around 20 V for producing amorphization. A main hurdle to the fabrication of this configuration is related to the ability of shaping ITO (lift-off). ITO can be properly engineered to have rather small values of $\kappa$ [24], thus reducing the overall optical insertion losses.

In the group of study, thanks to the superior $W$ heating capacitance, we achieve the shortest crystallization (~ 1 µs, 20 pulses) and Maximum temperature in the O-PCM layer as a function of time for rectangular pulse heating. amorphization pulse (~ 1 µs) period as well as the lowest threshold voltage (12 V). (**Fig.7, Tab.2**) This configuration has the drawback of introducing higher losses of ~0.1dB/µm due to the metal ohmic losses. Attention should be paid when implementing this electrothermal scheme when writing the memories with high bandwidth for avoiding carbonization of the GSST in proximity of the contact. A resistive heater optimized for efficient switching and contemporary not generating insertion losses, can be made in doped silicon or in silicide, currently used in p-n modulator, positioned next to the waveguide. [25]

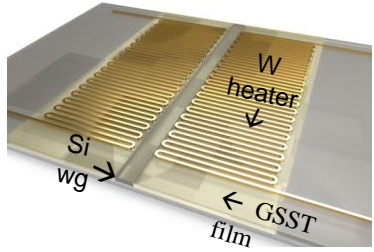

**Figure 3** 3D rendering of a lateral thermoelectric switching configuration. A heating serpentine made tungsten (W) is deposited on the side of a Si waveguide on top of a GSST film. The electrical current running through the W circuit dissipates energy in form of heat inducing a local phase transition.

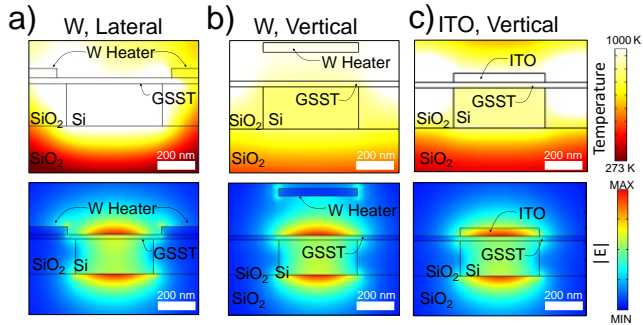

**Figure 6.** Numerical study (COMSOL) of the electro-thermal switching at the equilibrium (top row) and normalized electric field mode profile (bottom row) of hybrid Si-GSST waveguide. Heat map produced by Joule heating of a tungsten (a,b) and ITO (c) heating element in Vertical (a, c) and lateral (b) configuration. In the electrothermal simulation represented in figure a-c) the GSST is amorphous. Lateral configuration (a) provides the highest local heat in a shorter time with limited additional losses to the propagating mode (TM).

**Table 2** Comparison of the different heating element configurations for inducing thermoelectric switching. The amorphization temperature is considered 900 K and the crystallization temperature is considered 547 K (for 60 µs, 10 pulses).

| | Configuration | | |
|---|---|---|---|
| | **W, Lateral** | **W, Vertical** | **ITO, on top** |
| **Pulse duration (Crystallization)** | Multiple (20)pulses <1 µs | Multiple (20)pulses < 3 µs | Multiple (20)pulses < 3 µs |
| **Voltage (Crystallization)** | 5 V | | |
| **Pulse duration (Amorphization)** | ~ 1 µs | ~2 µs | < 2.5 µs |
| **Voltage (Amorphization)** | 12 V | 25 | 25 V |
| **Insertion Losses TM Mode [dB/µm]** | 0.11 dB/µm | 0.06 dB/µm | 0.14 dB/µm |

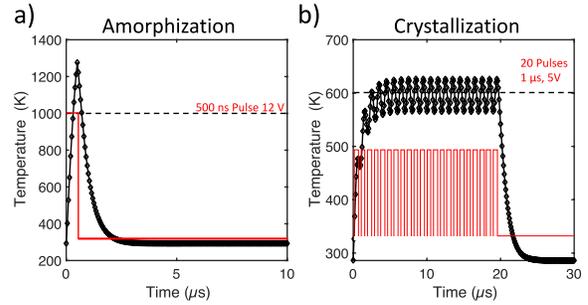

**Figure 7** Electro-thermal switching from crystalline to amorphous (a) and vice versa (b) through Joule heating in a lateral W configuration. Temperature average in the GSST layer as a function of time for rectangular pulse wave heating. The amorphization temperature is the melting temperature (>900K) while the temperature for crystallization (~523K but below amorphization temperature) is kept approximately constant for 20µs.

## D. Network
## In this section we trained g

The NN architecture that exploits the proposed non-volatile weighed addition can be emulated on an open source ML framework, i.e. Tensorflow. As preliminary study, we estimated the functionality of the proposed perceptron as main unit of the NN, by emulating its behavior in a 3-layer fully connected NN implemented in the Google Tensorflow tool and, as an initial example, for the MNIST data set, which is a well-known machine-learning data set comprised of 60,000 grayscale images of handwritten digits.

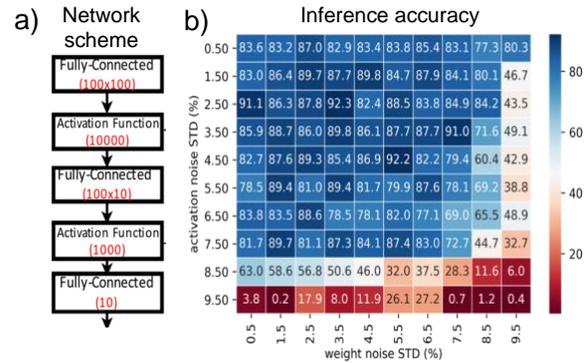

**Figure 8** a) Schematic of the fully connected network composed by 2 layers of 100 neurons. b) Accuracy results for the inference on unseen data for NN trained with 2% of noise.

The first layer does not perform any type of weighting to the inputs, while the second layer in our network also has 100 neurons, which receive inputs from the first layer with an all to all connection (100x100) and perform weighting (nonlinear quantized during inference). Nonlinear activation functions (here considered as electro-optic[26-30]) are placed between two consecutive layers on each input connection (**Fig. 8a**).

We note, that whole PCMs seem as an ideal material to provide the synaptic weights, other emerging materials for modulation co-integrated with silicon photonics could be used as well such as those based on ITO, Graphene, for example [28, 31,32]. In addition, novel modulation schemes could be explored as well that an modulation concepts such as attojoule efficient modulators [33-36].

For the training and inference tests here, we have 100x100 NLAFs operating between the first and the second layer. The third layer reduces the dimensionality of the network and comprises just 10 neurons. The network is trained both without and with noise of the weights and NLAF. Our hypothesis, confirmed in a recent publication[6] and from preliminary studies on the network is that, when we allow for a certain amount of noise during the training, the model during the inference stage, becomes more robust; the effect of adding a noise equivalent to 0.01% of the maximum signal swing at the output of neurons significantly improves inference, as shown in **Fig. 8b**. Note, the addition of this amount of noise during the training may result in small (~2%) accuracy loss for low level of noise during inference (**Fig. 8b**) . However, the model becomes more robust to higher levels of noise while performing inference. This shows that modeling noise by adding training noise can fine-tune the network for a physical noisy realization. Lastly adding further amount of noise beyond the initial 0.1% results in lower (<60%) inference accuracy. One important limitation of the proposed photonic NN based on PCM comes from the bit-resolution of the signal after weighting, which is quantized in just 32 states (considering a GSST layer with a lateral footprint of just 16 µm). Specifically, the weights in a first iteration of the network are restricted to be having 5-bit resolution. Additionally, in future works we will further study the effect of quantization of the weights while performing inference, since limiting the number of bits will simplify the complexity of the network implementation and reduce the number of states and consequently the overall footprint. We will also investigate the effect of 'pruning', by limiting the number of node-to-node connections to only the meaningful ones, aiming for reducing the network complexity without losing inference accuracy. To compensate the quantization error the solutions are manifolds and include; 1) increase the NN scale which provides greater expressive power, 2) adopt advanced quantization algorithms to better represent the information during training, and furthermore 3) gradually add the quantization constraints in a training-retraining flow which helps to converge to a better local optimum in the training.

## III. CONCLUSIONS AND OUTLOOK

In summary, we have investigated a low losses programmable Mach Zehnder modulator, based on a hybrid GSST-silicon waveguide, which showed a coherent quantized response as function of the portion of the phase change material that has been written by means of electrothermal switching. Furthermore, we modeled the transient thermal response of three distinct electrothermal heating configuration based on Joule heating. The photonic memory displays a quantized response of 3-bit in an extremely compact footprint of only 3.8µm, complemented by very small insertion losses, below 3dB, attributed primarily to the balancing mechanism for maximizing the extinction ratio. Moreover, the studied platforms provide insights into the speed of a photonic tensor processor architecture based on integrated photonic memories which stores the weights of a trained neural network and can be updated in parallel at sub MHz speed. The proposed optical module responses were used as weighting for a fully connected neural networks, emulated in Tensor Flow. We tested the quantized transfer function on a standardized neural network training set, MNIST classifiers of handwritten digits. Our results show that the neural network reaches very high level of accuracy in the inference phase and sufficiently robust. The simple and yet powerful architecture is a promising solution for optical information processing and vector matrix multiplication when performing inference, given that light is just attenuated (e.g. filtered) in a completely passive fashion and the operating speed of the entire network is limited only by the time of flight of the photon in the integrated platform. Also, each bit of the node has the potential to be intrinsically reconfigured, thence altering the weights and updating the neural network after successive trainings. Therefore, the proposed engine has the potential to significantly outperform in terms of computing speed and energy efficiency the established electronics or electro-optics technology. We envision that frequency selective memories based on GSST can improve the integrated all optical writing and the parallelism and consequently opening new frontiers in optical computing and communication [37], but could also be used in photonic analog comput2 systems such as partial differential equation solvers [38].

## ACKNOWLEDGMENT

V.S. is supported by the Presidential Early Career Award for Scientist and Engineers (PECASE) nominated by the Department of Defense (AFOSR).